\documentclass[floats,floatfix,amssymb,prl,superscriptaddress,twocolumn,aps]{revtex4}

\usepackage{graphicx, epsfig, amssymb} 
\usepackage{amsmath, amsfonts}
\usepackage{bm} 
\usepackage[caption=false]{subfig}
\usepackage[usenames]{color}
\usepackage[linktocpage]{hyperref}

\pdfoutput=1

\begin{document}

\title{Extremal Scalarization of Charged Black Holes: Miransky Scaling across Reissner–Nordstr\" om Extremality}

\author{Hong Guo}
\email{hong\_guo@pku.edu.cn}
\affiliation{Kavli Institute for Astronomy and Astrophysics, Peking University, Beijing 100871, China.}

\author{Yun Soo Myung}
\email{ysmyung@inje.ac.kr}
\affiliation{Center for Quantum Spacetime, Sogang University, Seoul 04107, Republic of Korea.}

\author{Lijing Shao}
\email{lshao@pku.edu.cn}
\affiliation{Kavli Institute for Astronomy and Astrophysics, Peking University, Beijing 100871, China.}
\affiliation{National Astronomical Observatories, Chinese Academy of Sciences, Beijing 100101, China.}

\begin{abstract}
We study spontaneous scalarization near and at extremality in Einstein–Maxwell–scalar theory.
As Reissner–Nordstr\"om (RN) extremality is approached from below, the nonextremal scalarization existence curves collapse toward a common critical threshold $\alpha_c=1/4$ given by the Breitenlohner-Freedman bound for the near-horizon throat (AdS$_2\times S^2$) of its extremal black hole. 
For a polynomial coupling, we construct extremal black holes with nonconstant scalar hair, satisfying $M^2+Q_s^2=Q^2$, which places them in the overcharged region $Q/M>1$ beyond the RN bound.
A common near-region phase mechanism predicts accumulation toward the RN extremality from both sides, yielding Miransky-type scaling. 
In the extremal sector, the logarithmic phase interval doubles, halving the exponent which denotes the geometric spacing of successive node scales.
\end{abstract}

\maketitle

\noindent{\bf {\em Introduction.}}
Spontaneous scalarization allows a black hole to develop scalar hair while the corresponding bald geometry remains an exact solution.
In Einstein-Maxwell-scalar (EMS) theory, a nonminimal scalar coupling to the Maxwell invariant leaves the Reissner–Nordstr\"om (RN) black hole solution intact but can trigger a tachyonic scalar mode.
The resulting scalarized nonextremal black holes form infinitely many branches labeled by the node number ($n=0,1,\cdots$)~\cite{Herdeiro:2018wub} and extend beyond the RN bound $q_e=1$.
Closely related mechanisms have already been found for curvature-induced scalarization in the Einstein-Gauss-Bonnet-scalar theory~\cite{Doneva:2017bvd,Silva:2017uqg,Antoniou:2017acq,Doneva:2022ewd}.
In EMS theory, the nonlinear branch structure depends sensitively on the coupling functions~\cite{Fernandes:2019rez,Blazquez-Salcedo:2020nhs,Guo:2025xwh}.
Extensions involving nonlinear electromagnetic interactions have also been investigated~\cite{Kiorpelidi:2023jjw}.
Despite this progress, the organization of the scalarized branches near RN extremality has remained unclear.

Let $q=Q/M$, with $q_e=1$ denoting RN extremality.
For the fundamental ($n=0$) existence curve, the limiting coupling $\alpha=1/4$ was already observed numerically~\cite{Herdeiro:2018wub}, but neither its geometric origin nor the accumulation of the excited curves ($n=1,\cdots$) was established.
The extremal limit is subtle because the RN near-horizon geometry becomes an infinitely long Bertotti--Robinson (BR) AdS$_2\times S^2$ throat~\cite{Bertotti:1959pf,Robinson:1959ev}.
Consequently, the asymptotic near-extremal behavior is difficult to infer from solutions at moderate charge.
An AdS$_2$ interpretation of an extremal scalarization threshold was noted in Einstein-Gauss-Bonnet-Maxwell-scalar theory~\cite{Brihaye:2019kvj}, where two branches for positive and negative Gauss-Bonnet couplings were obtained and a curvature singularity appeared before a regular scalarized extremal black hole was reached.

In this {\it Letter}, we examine specifically what happens when approaching to RN extremality during scalarziation in EMS theory. 
We first show that the nonextremal existence curves are organized by the Breitenlohner-Freedman (BF) bound of the near-horizon throat~\cite{Breitenlohner:1982jf}.
For RN, the BF bound is given by a critical threshold of $\alpha_c=1/4$ for the $s$-mode $(\ell=0)$ scalar.
Above this threshold, the throat scaling dimension becomes complex and the scalar oscillates along the logarithmic radial coordinate.
Phase matching then produces a geometric node sequence and the essential scaling $\sigma_n\sim\exp \big[-\mathrm{const.}/\sqrt{\alpha-\alpha_c} \big]$, characteristic of Berezinskii-Kosterlitz-Thouless (BKT)/Miransky-type scaling~\cite{Efimov:1970zz,Miransky:1984ef,Kaplan:2009kr,Jensen:2010ga,Iqbal:2010eh}. 
We also construct infinite many extremal black-hole branches with nonconstant scalar hair for a coupling $f(\phi)$ with a nonzero stationary point. 
A first integral $E$ shows that every such solution with scalar charge $Q_s\neq0$ and $f(\phi_\infty)=1$ lies at $q>q_e$, beyond the RN bound $q\le q_e$. 
The same near-region equation governs the RN zero modes for $q<q_e$ and the RN-like outer region of the nonlinear extremal solutions for $q>q_e$. 
On the extremal side, the available logarithmic phase interval doubles, changing the node spacing by a factor of two in the exponent.
Although the two families are physically distinct and not smooth continuations of one another, their near-extremal scaling is governed by the same throat dynamics.

\noindent{\bf {\em Model and its linearized theory.}}
We start with the EMS theory as
\begin{equation}
S=\frac{1}{16\pi}\int d^4x\sqrt{-g}\Big[R-2(\partial \phi)^2-f(\phi)F^2\Big]\label{action}	
\end{equation}
with two coupling functions, $f_1(\phi)=e^{\alpha\phi^2}$ and $f_2(\phi)=1+\alpha\phi^2-\beta\phi^4$, both satisfying $f(0)=1$, $f'(0)=0$ and $f''(0)=2\alpha$. 
The first two conditions preserve the RN solution, while the third imposes the tachyonic instability. 
With the ansatz $ds^2=-Ne^{-2\delta}dt^2+N^{-1}dr^2+r^2d\Omega_2^2$, $A=A_t(r)dt$ and $\phi=\phi(r)$, integrating the Maxwell equation gives $f(\phi)e^{\delta}r^2A_t'=Q$ and $F^2=-2Q^2/[f^2(\phi)r^4]$.  
The complete field equations are given in the Supplemental Material~\cite{SM}.

Linearizing around the RN background, the scalar perturbation satisfies $(\bar{\square}-\mu^2)\delta \phi=0$, with mass $\mu^2=\frac{f''(0)}{4}\bar{F}^2=-\frac{\alpha Q^2}{r^4}$. 
We use barred notation to denote background quantities and operators.
For $\alpha>0$, this mass is negative, yielding a tachyonic scalar mode.
Since both coupling functions have the same quadratic term, they lead to the same linearized equation. 
For fixed $q<q_e$, horizon regularity and decay at infinity define an eigenvalue problem for the coupling constant $\alpha$~\cite{Herdeiro:2018wub,Myung:2019oua,Hod:2020cal}. 
Its discrete eigenvalues $\alpha_n(q)$ are labeled by the node number $n$, with each eigenmode identifying the onset of the corresponding scalarized branch.


\noindent{\bf {\em The critical throat and nonextremal side.}}
The near-horizon geometry of an extremal RN black hole is the BR spacetime AdS$_2\times S^2$~\cite{Bertotti:1959pf,Robinson:1959ev}, with squared radii $v_0=1/N_2$ and $v_1=r_e^2$ where $N_2=\bar N''(r_e)/2$. 
Stability of the $s$-mode scalar requires the AdS$_2$ BF bound $m^2_{\mathrm{AdS}_2}v_0\ge-1/4$~\cite{Breitenlohner:1982jf}.
In this case, the critical threshold takes the form
\begin{equation}
\alpha_c=\frac{r_e^4N_2}{4Q^2},
\label{eq:alphac}
\end{equation}
which leads to $\alpha_c=1/4$ for RN black holes. 
\begin{figure}[t]
\begin{center}
\includegraphics[width=0.35\textwidth]{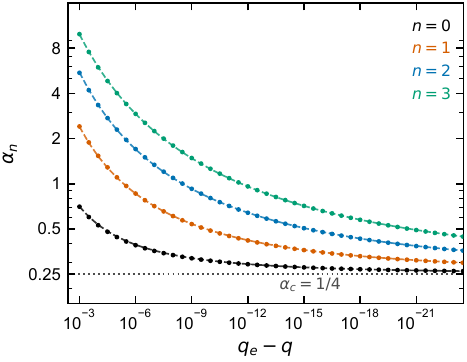}
\caption{Four lowest branches $\alpha_n$ approach the BF threshold $\alpha_c=1/4$ as $q_e-q\to0$. 
Dots denote the numerical eigenvalues, while dashed curves show the accumulation law in Eq.~\eqref{eq:collapse} with a branch-independent finite-throat phase offset determined by a single global fit, as detailed in the Supplemental Material~\cite{SM}. 
The dotted line marks the critical threshold $\alpha_c=1/4$ from Eq.~(\ref{eq:alphac}), toward which all four branches collapse.}
\label{fig:collapse}
\end{center}
\end{figure}
For examples, consider the four lowest existence curves as $q\to q_e$. 
Their asymptotic approach is described by
\begin{equation}
\alpha_n-\alpha_c\simeq\frac{4\alpha_c\pi^2(n+n_0^-)^2}{\hat L^2},
\hat L=\ln\frac{2M}{\sigma}, \sigma=M\sqrt{|1-q^2|},
\label{eq:collapse}
\end{equation}
where $n_0^-$ is an $O(1)$ phase offset and $\hat L$ is the logarithmic throat length, which diverges at extremality.
For any node number $n$, Eq.~\eqref{eq:collapse} gives $\alpha_n-\alpha_c\propto\hat L^{-2}$. 
This implies that the existence curves approach the BF threshold and their spacing shrinks to zero as $q\to q_e$. 
Note that the convergence is slow because $\hat L$ grows logarithmically.
Hence, a branch can remain visibly above $\alpha=\alpha_c$ even it is extremely close to extremality. 
Figure~\ref{fig:collapse} indicates the asymptotic scaling in Eq.~\eqref{eq:collapse}, with $n_0^-$ obtained from a single global fit to all numerical eigenvalues. 
The four branches therefore collapse at the common threshold $\alpha_c=1/4$. 

The same relation also determines how the spectrum is organized for $\alpha>\alpha_c$. 
Defining $\nu=\tfrac12\sqrt{(\alpha-\alpha_c)/\alpha_c}$, inverting Eq.~\eqref{eq:collapse} gives $\hat L_n\simeq\pi(n+n_0^-)/\nu$.
Increasing $n$ by one adds a phase $\pi$ and therefore increases the required logarithmic throat length by $\pi/\nu$.
Considering $\sigma=2M e^{-\hat L}$, the onset scales form the geometric sequence $\sigma_{n+1}/\sigma_n\simeq e^{-\pi/\nu}$.
The number of admissible modes below a fixed coupling grows as $\mathcal{N}\sim\hat L\nu/\pi$. 
Thus, finitely many modes fit within a throat with finite length, while an unbounded sequence can accumulate as $\hat L\to\infty$. 
The exponential generation of scales together with the geometric spectrum is the characteristic Miransky structure~\cite{Miransky:1984ef}.

\noindent{\bf {\em Scalarized extremal black holes.}}
At extremality, the field equations impose
\begin{equation}\label{eq:hor}
r_+^2=\frac{Q^2}{f(\phi_h)},\qquad f'(\phi_h)=0,\qquad N_2=\frac{1}{r_+^2}.
\end{equation}
An integer-power expansion admits only a constant-scalar solution. 
For a coupling $f_1$, whose stationary point is $\phi=0$, this corresponds to the bald extremal configuration. 
But the polynomial coupling $f_2$ has the nonzero stationary point $\phi_h=\sqrt{\alpha/(2\beta)}$ with $f(\phi_h)=1+\alpha^2/(4\beta)$.
This changes the local extremal horizon data without modifying the linearized scalar equation. 
The constant-scalar-hair extremal solution has  $\phi_\infty=\phi_h$ and $Q_s=0$, and is therefore distinct from the asymptotically flat nonconstant scalar family below~\cite{SM}.

\begin{figure}[t!]
\centering
\includegraphics[width=\columnwidth]{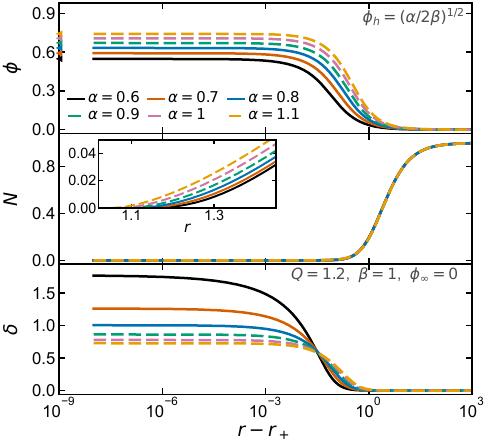}
\caption{Profiles of scalarized extremal black holes with nonconstant scalar hair for $Q=1.2$, $\beta=1$, $\phi_\infty=0$ with coupling constants  $\alpha=0.6, \, 0.7, \, 0.8, \, 0.9, \, 1.0, \, 1.1$.
({\it Left}) Scalar hair $\phi(r-r_+)$.
({\it Middle}) Metric function $N(r-r_+)$. 
({\it Right}) Redshift metric function $\delta(r-r_+)$. 
}
\label{fig:profiles}
\end{figure}

The degenerate horizon represents a regular singular point.
A Frobenius branch of $\phi=\phi_h+c\,\rho^{\lambda_+}+\cdots$ survives after discarding the divergent local solution.
Here, $c$ is a free coefficient that serves as the shooting parameter, connecting the nonzero horizon value $\phi_h$ to $\phi_\infty=0$.
The indicial equation is given by $\lambda(\lambda+1)=m_h^2v_0$ with $m_h^2v_0=2\alpha/[1+\alpha^2/(4\beta)]$.
Since the regular branch is nonanalytic in $\rho=r-r_+$ for generic $\lambda_+$, an ordinary Taylor expansion removes this degree of freedom and leaves only constant scalar hair. 
The Frobenius mode is thus essential for a nonconstant scalar profile to emerge from the degenerate horizon.
Requiring the scalar backreaction to remain subleading relative to the quadratic zero of $N$ further gives $\lambda_+>1/2$.
For the $f_2(\phi)$ coupling, this condition becomes $m^2v_0=2\alpha/(1+\alpha^2/4\beta)>3/4$. 
At $\beta=1$, it gives $0.3892<\alpha<10.2775$ and we denote the lower bound by $\alpha_*=0.3892$.
This bound is necessary for the Frobenius construction, but the global existence still requires a successful shooting to infinity. 
Scalarized extremal solutions with nonconstant scalar hair are constructed numerically and their profiles are displayed as functions of $r-r_+$ in Fig.~\ref{fig:profiles}.

We note that the above construction also explains the relation to earlier results. 
For $f_1$, scalarization can extend into the overcharged region, but Eq.~\eqref{eq:hor} forces $\phi_h=0$, excluding the nonconstant extremal branch considered here~\cite{Astefanesei:2019pfq}.
Additional gauge fields can modify the horizon balance and allow scalarized extremal solutions~\cite{Chew:2026clh,Chen:2026olq}.
Here, this role is replaced by the nonzero stationary point of $f_2$. 
The relevant ingredient is therefore the structure of $f$ at the degenerate horizon, rather than additional electric charges.

\noindent{\bf {\em Scalar charge at extremality.}}
Shooting from the degenerate horizon may yield a sequence of extremal solutions distinguished by the number of scalar nodes.
At fixed $Q$, $\alpha$, and $\beta$, these solutions share the same local horizon data and the same entropy, while their exterior profiles and asymptotic charges differ. 
A central constraint is provided by the radial first integral
\begin{equation}
E\equiv e^{-2\delta}\bigg[\Big(\tfrac12r^2N'-r^2N\delta'\Big)^2+r^4N^2\phi'^2-\frac{NQ^2}{f(\phi)}\bigg],
\label{eq:firstint}
\end{equation}
which satisfies $E'=0$ independently of the specific form of $f(\phi)$.
A symbolic proof and numerical checks are carried out in the Supplemental Material~\cite{SM}.
It corresponds to the static one-dimensional Hamiltonian constraint with effective potential $Q^2/f(\phi)$~\cite{Ferrara:1997tw}.
Evaluating it at the horizon and at infinity gives a relation 
\begin{equation}
M^2+Q_s^2=\frac{Q^2}{f(\phi_\infty)}+\big(2T_HS\big)^2,
\label{eq:budget}
\end{equation}
which links the ADM mass, electric charge, scalar charge, temperature, and entropy without numerical interpolation between the two boundaries. 
The first integral shows how the asymptotic scalar charge $Q_s$ enters the extremality relation.
For $T_H=0$, $\phi_\infty=0$ and $f(0)=1$, one finds 
\begin{equation}
M^2+Q_s^2=Q^2.
\label{eq:extrel}
\end{equation}
Consequently, every solution with $Q_s\neq0$ satisfies
\begin{equation}
q=\frac{1}{\sqrt{1-Q_s^2/Q^2}}>1.
\label{eq:over}
\end{equation} 
Here, ``overcharge'' refers to $q>q_e$ relative to the bald RN bound, while the scalarized solution retains a regular extremal horizon. 
Thus $q>1$ is maintained by the nonzero scalar charge. 
For small $Q_s$, $q-1=\frac{Q_s^2}{2Q^2}+O(Q_s^4)$, so the shift from RN extremality starts quadratically in $Q_s^2$.
The extremal family thus represents a $T_H=0$ subset of the broader overcharged scalarized sector.

Equation~\eqref{eq:extrel} identifies the near-region scale exactly as $\sigma=M\sqrt{q^2-1}=|Q_s|$.
For these solution, the Komar mass is $M=2T_HS+\Phi Q$, with gauge potential $\Phi=A_t(\infty)-A_t(r_+)$. 
At extremality, this reduces to $M=\Phi Q$. 
Combining it with Eq.~\eqref{eq:extrel} leads to
\begin{equation}
\Phi^2+\Big(\frac{Q_s}{Q}\Big)^2=1,
\label{eq:unit}
\end{equation}
which relates the near-region scale, the asymptotic scalar charge, and the gauge potential through an exact identity. 

\begin{figure}[t]
\centering
\includegraphics[width=0.82\columnwidth]{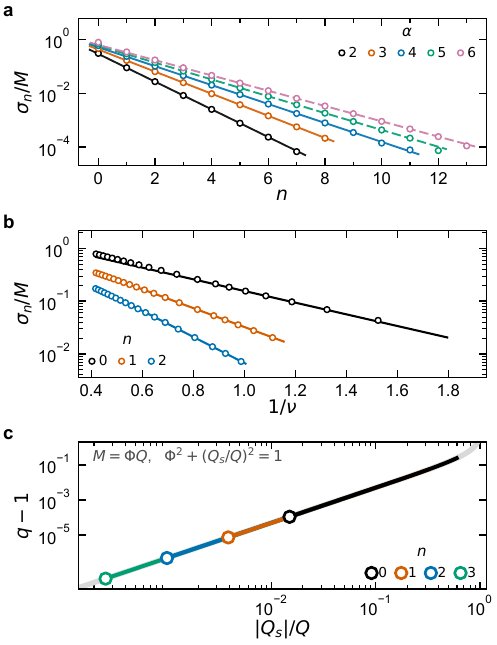}
\caption{Scalaized extremal solutions at $\beta=1$ and $Q=1.2$. 
(a) At fixed $\alpha$, $\sigma_n/M=|Q_{s,n}|/M$ forms a geometric node sequence. 
Lines are described by $\sigma_n\simeq2M\exp\left[-(n+n_0^+)\pi/2\nu\right]$, with the slope in $n$ fixed by $\nu$, and the intercept by $n_0^+$ measured on each family. 
(b) Fixed-node branches show the essential dependence on $1/\nu$. 
(c) Overcharge from the gauge potential agrees with the exact scalar-charge relation in Eq.~\eqref{eq:unit} for the common domain $n\leq3$ and $\alpha\geq1.5$. 
Open circles locate the lowest scalar charge reached on each branch.}
\label{fig:bkt}
\end{figure}

\noindent{\bf {\em Scalaized extremal family.}}
Figure~\ref{fig:bkt} summarizes the scaling structure of the scalarized extremal family. 
The quantity $\sigma_n/M$ in panels (a) and (b) denotes the normalized scalar charge of the $n$-node extremal solution. 
Panel (a) indicates the node-number scaling at fixed coupling.
For each $\alpha$, $\ln \big[|Q_s|/M \big]$ varies  linearly with the node number $n$, showing that the scalar charges form a geometric sequence as 
\begin{equation}
\frac{|Q_{s,n+1}|}{|Q_{s,n}|}\simeq e^{-\pi/(2\nu)}.
\label{eq:extratio}
\end{equation}
Equivalently, one has a linear dependence of $\ln[\sigma_n/M]$ on $n$ with slope $-\pi/(2\nu)$. 
This slope is fixed by $\nu$ and is not fitted independently for each sequence. 
Compared with the nonextremal onset relation, the logarithmic spacing is reduced by a factor of two. 
Panel (b) shows each $n$-node branch as $\alpha$ varies.
Considering $\nu\propto\sqrt{\alpha-\alpha_c}$, the throat prediction has the essential Miransky dependence $\ln[\sigma_n/M]\propto-1/\nu$ rather than a power law in $\alpha-\alpha_c$. 
The numerical extremal branches show this functional form over the accessible interval, while the present Frobenius construction terminates at $\alpha_*>\alpha_c$. 
Panels (a) and (b) probe complementary consequences of the same quantization condition. 
Panel (c) tests the exact charge relation using independently extracted quantities on the two axes.
The horizontal coordinate $|Q_s|/Q$ comes from the asymptotic scalar charge, whereas the vertical one follows from the gauge potential.
At extremality, $M=\Phi Q$ gives $q-1=(1-\Phi)/\Phi$, while Eq.~\eqref{eq:unit} predicts $q-1=[1-(Q_s/Q)^2]^{-1/2}-1$.
The numerical data agree with this relation, providing a cross-sector consistency check between independently extracted scalar and gauge potential. 
For small $Q_s$, Eq.~\eqref{eq:over} shows that the spacing $e^{-\pi/(2\nu)}$ in scalar charge becomes $e^{-\pi/\nu}$ in $q-1$.

\noindent{\bf {\em Two-sided accumulation.}}
The results reveal two physically distinct sequences converging toward the RN extremality at $q=q_e$.
For $q\le q_e$, the numerically determined nonextremal existence curves approach $\alpha_c=1/4$ as $q\to q_e^-$. 
For $q>q_e$, decreasing $\alpha$ sends $Q_s\to0$ and thus $q\to q_e^+$.
The first sequence consists of linear scalar modes on nonextremal RN, whereas the second consists of fully nonlinear scalars with $T_H=0$. 
They approach $q_e$ from opposite sides without forming a smooth continuation through $q=q_e$.
Instead, RN extremality acts as their common asymptotic accumulation point under the same near-throat dynamics.

\begin{figure}[t!]
\centering
\includegraphics[width=0.35\textwidth]{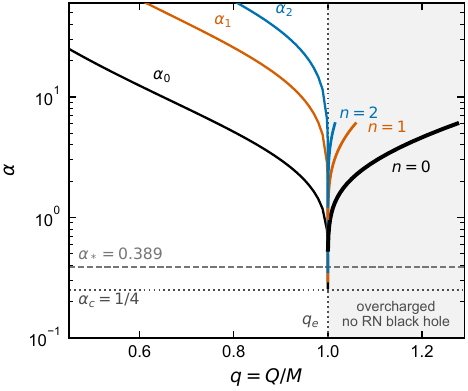}
\caption{Two-sided scalarization structure for $\beta=1$ in the $(q,\alpha)$ plane. 
Colored thin curves for $q\leq q_e$ represent the $n=0,1,2$ existence curves $\alpha_n(q)$, while thick curves denote the $T_H=0$ extremal branches for $q>q_e$.
The two families meet on the dotted vertical line at RN extremal point $q=q_e=1$ from opposite sides.}
\label{fig:qalpha}
\end{figure}

The scales $\alpha_c$ and $\alpha_*$ have distinct origins. 
The former is the BF threshold of the RN throat, whereas the latter marks the breakdown of the Frobenius hierarchy used here.
Thus, the interval $\alpha_c<\alpha<\alpha_*$ remains unresolved by the present construction. 
Figure~\ref{fig:qalpha} displays this difference.
The two convergence behaviors can be traced to a single near-region equation, which also accounts for the factor of two in the node spacing.
As $Q_s\to0$, the exterior outside the near-horizon scalarized region approaches an RN-like region with the same asymptotic $M$ and $Q$.
Introducing $y=(r-M)/\sigma$, the $s$-wave equation reduces at leading order to
\begin{equation}\label{eq:bfv}
\frac{d}{dy}\Big[(y^2\mp1)\frac{d\phi}{dy}\Big]=m^2v_0\,\phi,
\qquad m^2v_0=-\frac{\alpha}{4\alpha_c},
\end{equation}
where the minus and plus signs correspond to below and above RN extremality, respectively. 
At large $|y|$, the exponents are $-1/2\pm i\nu$, so that the scalar oscillates along  $\ln|y|$ with frequency $\nu$ and accumulates a phase proportional to the logarithmic throat length. 
Below extremality, $y=1$ is the outer horizon. Choosing $y=\cosh s$ and $\phi=u/\sqrt{\sinh s}$, the near region is a one-sided interval of length $\hat L+O(1)$.  
Above extremality, $y^2+1$ has no real zero in the RN region. 
With $y=\sinh t$ and $\phi=u/\sqrt{\cosh t}$, the solution traverses both sides of the neck, giving a total interval $2\hat L+O(1)$. 
The $O(1)$ endpoint shifts affect the phase offset only. 
The corresponding reduced equations take the forms
\begin{equation}
u''+\Big[\nu^2+\tfrac14\,\mathrm{csch}^2s\Big]u=0,
u''+\Big[\nu^2-\tfrac14\,\mathrm{sech}^2t\Big]u=0,
\label{eq:schro}
\end{equation}
referring to $q<q_e$ and $q>q_e$, respectively.
Here, the available logarithmic phase interval is $\nu\hat L$ below extremality and $2\nu\hat L$ above it, yielding the phase quantization $\Theta_n\equiv\frac{\nu\hat L_n}{\pi}=n+n_0^-$ for $q<q_e$ and $\Theta_n\equiv\frac{2\nu\hat L_n}{\pi}=n+n_0^+$ for $q>q_e$.
Since $\nu\propto\sqrt{\alpha-\alpha_c}$, the exponential dependence on $1/\nu$ gives Miransky rather than power-law scaling in $\alpha-\alpha_c$.

On the nonextremal side, the exact scalar-cloud solution is a Legendre function. 
Imposing its zero at infinity and using the large-argument connection formula give $n_0^-=1-(2\ln2/\pi)\nu+O(\nu^2)$ and $\sigma_n\simeq8M\exp[-(n+1)\pi/\nu]$ near the BF critical threshold.
On the overcharged side, the near-region equation fixes the logarithmic spacing in Eq.~\eqref{eq:extratio}, whereas $n_0^+$ depends on the near-horizon scalarized region and must be determined from the full scalarzied extremal solution.
Therefore, the same BF-controlled frequency $\nu$ governs the near-region scaling on both sides of extremality, even though the two families have different boundary conditions and nonlinear completions.

\noindent{\bf {\em Conclusions.}}
In EMS theory, the extremal RN AdS$_2\times S^2$ throat provided the organizing principle for the extremal scalarization.
Its BF bound determined the common accumulation threshold $\alpha_c=1/4$ for the $s$-mode scalar clouds.
On the nonextremal side, the logarithmic throat phase quantized the RN zero modes and yields both the geometric node sequence and Miransky-type essential scaling.

The nonlinear extremal sector is physically distinct.
A nonconstant scalar hair solution with $T_H=0$ requires a nonzero stationary point of $f_2(\phi)$ and a regular Frobenius branch at the degenerate horizon.
For an asymptotically flat branch, the exact first integral gives $M^2+Q_s^2=Q^2$.
Hence, $Q_s\neq0$ shifts the extremality condition to $q>1$ while preserving a regular degenerate horizon, so $Q_s$ measures the displacement from RN extremality rather than a loss of black-hole regularity.

The same BF-controlled near-region equation governs the RN-like outer region of the extremal solutions for $q>q_e$.
Below extremality, the oscillatory region is one-sided, whereas above extremality, it extends across both sides of the long throat before reaching the near-horizon scalarized region.
The doubled phase interval changes the geometric spacing from $e^{-\pi/\nu}$ to $e^{-\pi/(2\nu)}$.
The nonlinear interior shifts the phase offset but not this asymptotic spacing, so two physically distinct families are organized by the same near-throat dynamics.

The interval $\alpha_c<\alpha<\alpha_*$ remains unresolved because the Frobenius hierarchy used breaks down at $\alpha_*$. 
This limitation does not exclude solutions in that interval.
Determining the nonlinear phase analytically and testing the same BF-controlled mechanism in other scalarized charged black holes are natural next steps.

\bigskip

\noindent{\bf {\em Acknowledgements.}}
We appreciate Stoytcho Yazadjiev and Daniela Doneva for their helpful discussion. 
Y.S.M. is supported by the National Research Foundation of Korea (NRF) grant funded by the Korea government (MSIT) (RS-2022-NR069013). 
L.S.\ is supported by the Max Planck Partner Group Program funded by the Max Planck Society.


\clearpage

\onecolumngrid

\setcounter{section}{0}
\renewcommand{\thesection}{\arabic{section}}
\setcounter{equation}{0}
\renewcommand{\theequation}{S\arabic{equation}}
\setcounter{figure}{0}
\renewcommand{\thefigure}{S\arabic{figure}}
\setcounter{table}{0}
\renewcommand{\thetable}{S\arabic{table}}
\begin{center}
\Large\bfseries Supplementary Materials
\end{center}

\section{Field equations and the linear scalar-cloud problem}
\label{sec:eom}

In this Einstein-Maxwell-scalar theory, the equations of motion are
\begin{align}
m'&=\frac12\left[r^2N\phi'^2+\frac{Q^2}{f(\phi)r^2}\right],
\label{eq:m}\\[2pt]
\delta'&=-r\phi'^2,
\label{eq:delta}\\[2pt]
\left(e^{-\delta}r^2N\phi'\right)'&=-\frac{f_{,\phi}}{2f^2}\frac{Q^2e^{-\delta}}{r^2}.
\label{eq:scalar}
\end{align}
Here $f_{,\phi}=df/d\phi$. 
Eliminating $m$ in favor of $N$ gives
\begin{equation}
N'=\frac{1-N}{r}-rN\phi'^2-\frac{Q^2}{r^3f(\phi)}.
\label{eq:Nprime}
\end{equation}
At large $r$, an asymptotically flat solution behaves as
\begin{equation}
N=1-\frac{2M}{r}+\frac{Q_s^2+Q^2}{r^2}+O(r^{-3}),
\qquad
\phi=\phi_\infty+\frac{Q_s}{r}+O(r^{-2}),
\label{eq:asymptotics}
\end{equation}
with $\delta=Q_s^2/(2r^2)+O(r^{-3})$, $M$ and $Q_s$ denote the ADM mass and scalar charge. 
The scalarized branch studied in the Letter has $\phi_\infty=0$ and hence $f(\phi_\infty)=1$.
Although the two couplings agree to quadratic order around $\phi=0$, their nonlinear extremal sectors differ because a regular degenerate horizon probes the stationary points of $f$.
We keep $\beta$ explicit below and set $\beta=1$ for comparison with the numerical solutions.

Linearizing Eq.~\eqref{eq:scalar} around RN, with $\bar N=1-2M/r+Q^2/r^2=(r-r_+)(r-r_-)/r^2$, gives the $\ell=0$ scalar equation
\begin{equation}
\left(r^2\bar N\phi'\right)'+\frac{\alpha Q^2}{r^2}\phi=0.
\label{eq:linear}
\end{equation}
At fixed $q\equiv Q/M<1$, regularity at the outer horizon removes the logarithmically divergent local solution, while $\phi_\infty=0$ selects the decaying solution at infinity. 
These two boundary conditions select the discrete eigenvalues $\alpha_n(q)$ associated with the scalar-cloud branches.
At extremal RN, $r_e=M=Q$ and $\bar N=N_2(r-r_e)^2+\cdots$ with $N_2=1/r_e^2$. 
The near-horizon geometry is AdS$_2\times S^2$ with AdS$_2$ radius squared $v_0=1/N_2$. 
In this region, the scalar equation takes the form of a massive scalar on AdS$_2$, with
\begin{equation}
m^2v_0=-\frac{\alpha Q^2}{r_e^4N_2}
=-\frac{\alpha}{4\alpha_c},
\qquad
\alpha_c\equiv\frac{r_e^4N_2}{4Q^2}.
\end{equation}
The AdS$_2$ BF bound $m^2v_0\ge-1/4$ therefore becomes $\alpha\le\alpha_c$~\cite{Breitenlohner:1982jf} with $\alpha_c=1/4$ for extremal RN. 

\section{Extremal scalarized solutions}
\label{sec:extremal}

Let $\rho=r-r_+$ and consider a regular degenerate horizon with $N=N_2\rho^2+\cdots$ and finite $\delta_h$. 
Setting $N(r_+)=N'(r_+)=0$ in Eq.~\eqref{eq:Nprime} gives
$r_+^2=Q^2/f(\phi_h)$. 
Since $N\sim\rho^2$, the left-hand side of Eq.~\eqref{eq:scalar} vanishes at the horizon, requiring $f_{,\phi}(\phi_h)=0$. 
For the Frobenius class considered below, the scalar backreaction is subleading to the quadratic zero of $N$. 
Differentiating Eq.~\eqref{eq:Nprime} at the horizon then gives $N''(r_+)=2/r_+^2$, or $N_2=1/r_+^2$. 
These relations reproduce Eq.~\eqref{eq:hor}.

An ordinary Taylor expansion about a degenerate horizon admits only a constant scalar hair. 
For the nonzero stationary point $\phi_h=\sqrt{\alpha/(2\beta)}$ of the polynomial coupling $f_2$, the constant-scalar geometry takes the extremal RN form, giving
\begin{equation}
r_+=\frac{Q}{\sqrt{1+\alpha^2/(4\beta)}},
\qquad
S=\frac{\pi Q^2}{1+\alpha^2/(4\beta)},
\qquad
q\equiv\frac{Q}{M}=\frac{Q}{r_+}=\sqrt{1+\frac{\alpha^2}{4\beta}}.
\end{equation}
This constant hair solution has $\phi_\infty=\phi_h$ and $Q_s=0$, and is therefore distinct from the nonconstant branch with $\phi_\infty=0$ studied in the Letter.
The same local horizon relations can also be recovered from the entropy-function construction on AdS$_2\times S^2$~\cite{Sen:2005wa}.

A nonconstant family is possible because the degenerate horizon is a regular singular point. 
We therefore allow the Frobenius correction
\begin{equation}
\phi=\phi_h+c\rho^\lambda+\cdots.
\label{eq:frob}
\end{equation}
Since $f_{,\phi}=f_{,\phi\phi}(\phi_h)c\rho^\lambda+\cdots$, the leading terms in Eq.~\eqref{eq:scalar} give
\begin{equation}
\lambda(\lambda+1)=m_h^2v_0,
\qquad
m_h^2=-\frac{f_{,\phi\phi}(\phi_h)Q^2}{2f^2(\phi_h)r_+^4},
\qquad
v_0=\frac{1}{N_2}=r_+^2.
\label{eq:resonance}
\end{equation}
Thus the exponent is fixed by the local AdS$_2$ throat and the curvature of the coupling function at its nonzero stationary point. 
The regular root is
\begin{equation}
\lambda_+=\frac{-1+\sqrt{1+4m_h^2v_0}}{2},
\label{eq:lambdaplus}
\end{equation}
while $\lambda_-=-1-\lambda_+<-1$ diverges.
The coefficient $c$ remains free locally and is fixed by shooting to the asymptotic condition $\phi_\infty=0$.

Regularity further requires $\lambda_+>1/2$.  
Equation~\eqref{eq:delta} gives $\delta-\delta_h\sim\rho^{2\lambda-1}$, while the scalar contribution to Eq.~\eqref{eq:Nprime} produces $\Delta N\sim\rho^{2\lambda+1}$.
Requiring finite $\delta_h$ and $\Delta N=o(\rho^2)$ therefore gives $\lambda_+>1/2$, equivalently $m_h^2v_0>3/4$. 
For $f_2$, this becomes $3\alpha^2-32\beta\alpha+12\beta<0$, or
\begin{equation}
\frac{16\beta-\sqrt{256\beta^2-36\beta}}{3}
<\alpha<
\frac{16\beta+\sqrt{256\beta^2-36\beta}}{3}.
\label{eq:window}
\end{equation}
A real interval exists for $\beta>9/64$. 
For $\beta=1$, this gives $0.389201<\alpha<10.277466$, with the lower endpoint $\alpha_*\simeq0.3892$ used in the Letter.
This is a local regularity condition and is distinct from the RN BF threshold $\alpha_c=1/4$.
The latter is determined by perturbations about the RN throat at
$\phi=0$, whereas $\alpha_*$ follows from Frobenius regularity about the nonzero stationary point $\phi_h$.

\section{Radial first integral and the extremality relation}
\label{sec:first}

The charge relation used in the Letter follows from a radial first integral that is independent of the specific form of $f(\phi)$. 
Define
\begin{align}\label{eq:first}
 K&\equiv\frac12r^2N'-r^2N\delta',\\[-2pt]
 E&\equiv e^{-2\delta}\left[K^2+r^4N^2\phi'^2-\frac{NQ^2}{f}\right].
\end{align}
Differentiating $E$ gives
\begin{align}
\frac12e^{2\delta}E'={}&K\left(K'-\delta'K\right)+r^4N^2\phi' \left[\phi''+\left(\frac2r+\frac{N'}N-\delta'\right)\phi'\right]-\frac{Q^2}{2f}\left[N'-2N\delta'-N\frac{f_{,\phi}}{f}\phi'\right].
\label{eq:Eprime1}
\end{align}
Expanding Eq.~\eqref{eq:scalar} gives $\phi''+\left(\frac2r+\frac{N'}N-\delta'\right)\phi'=-\frac{f_{,\phi}Q^2}{2f^2r^4N}$, while differentiating $K$ and using Eqs.~\eqref{eq:Nprime} and \eqref{eq:delta} yields $K'-\delta'K=\frac{Q^2}{fr^2}$.
Substituting into Eq.~\eqref{eq:Eprime1} gives
\begin{align}
\frac12e^{2\delta}E'
&=\frac{Q^2}{f}
\left[\frac{K}{r^2}-\frac12\left(N'-2N\delta'\right)\right]
=0,
\end{align}
where the last equality follows from the definition of $K$. 
Hence $E$ is radially conserved.

Using Eq.~\eqref{eq:asymptotics} and
$\delta=Q_s^2/(2r^2)+\cdots$, the conserved quantity at infinity is $E_\infty=M^2+Q_s^2-Q^2/f_\infty$.
At a regular horizon, $N(r_+)=0$ and $K_h=r_+^2N'(r_+)/2$, so one obtains
\begin{equation}
E_h=e^{-2\delta_h}K_h^2=\left[\frac{r_+^2N'(r_+)e^{-\delta_h}}{2}\right]^2=(2T_HS)^2,\nonumber
\end{equation}
where $T_H=N'(r_+)e^{-\delta_h}/(4\pi)$ and $S=\pi r_+^2$. 
Equating the two boundary values reproduces the general identity
in Eq.~\eqref{eq:budget}.
For bald RN, $Q_s=0$, $f=1$, and $2T_HS=\sqrt{M^2-Q^2}$, reducing Eq.~\eqref{eq:budget} to the usual RN relation. 

For the timelike Killing vector $\partial_t$, the Komar mass on a sphere of radius $r$ is $M_K(r)=\frac{r^2}{2}e^{-\delta}\left(N'-2N\delta'\right)=e^{-\delta}K$.
Using $K'-\delta'K=Q^2/(fr^2)$ obtained above, one has $\frac{dM_K}{dr}=e^{-\delta}\left(K'-\delta'K\right)=\frac{Q^2e^{-\delta}}{fr^2}=Q A_t'$. 
As $r\to\infty$, $M_K\to M$. 
At the horizon, $M_K(r_+)=r_+^2N'(r_+)e^{-\delta_h}/2=2T_HS$. 
Therefore, integrating the last equation from $r_+$ to infinity gives
\begin{equation}
M=2T_HS+\Phi Q,
\qquad
\Phi\equiv A_t(\infty)-A_t(r_+)
=\int_{r_+}^{\infty}\frac{Qe^{-\delta}}{f(\phi)r^2}\,dr.
\label{eq:smarr}
\end{equation}
At extremality, $T_H=0$ and hence $M=\Phi Q$. 
Combining this result with Eq.~\eqref{eq:extrel} reproduces Eq.~\eqref{eq:unit}.
The two terms in this relation come from different sectors of the solution. 

\section{Near-extremal throat and phase quantization}
\label{sec:phase}

The same BF-violating throat equation governs both sides of RN
extremality, while the corresponding radial domains are different.
For a near-extremal RN solution, $\sigma=\sqrt{M^2-Q^2}$ gives $r^2\bar N=(r-M)^2-\sigma^2=\sigma^2(y^2-1)$. 
For a weakly scalarized extremal solution, Eq.~\eqref{eq:extrel} instead gives $\sigma=\sqrt{Q^2-M^2}=|Q_s|$. 
Outside the near-horizon scalarized region, the geometry approaches the RN-like form with the same $M$ and $Q$, for which the corresponding leading expression is $r^2N\simeq\sigma^2(y^2+1)$. 
We therefore use the common notation
\begin{equation}
\sigma\equiv\sqrt{|M^2-Q^2|}=M\sqrt{|1-q^2|},
\qquad
y\equiv\frac{r-M}{\sigma}.
\label{eq:ydef}
\end{equation}

In the overlap region $\sigma\ll |r-M|\ll M$, where $r\simeq Q\simeq M$, Eq.~\eqref{eq:linear} reduces on both sides to Eq.~\eqref{eq:bfv}.
For $|y|\gg1$, a power law $\phi\sim|y|^\lambda$ gives $\lambda(\lambda+1)=m^2v_0$. 
Above the BF threshold,
\begin{equation}
\lambda_\pm=-\frac12\pm i\nu,
\qquad
\nu\equiv\sqrt{-\frac14-m^2v_0}
=\frac12\sqrt{\frac{\alpha-\alpha_c}{\alpha_c}}.
\label{eq:nu}
\end{equation}
The corresponding real solutions behave as
\begin{equation}
\phi\sim |y|^{-1/2}\cos\!\left(\nu\ln|y|+\vartheta\right).
\label{eq:logosc}
\end{equation}
The BF violation thus turns the real AdS$_2$ scaling dimension into a complex pair. 
Radial evolution through the throat becomes an oscillation in $\ln|y|$, with each additional node contributing approximately one extra phase interval $\pi$.

The difference between the two spectra follows from the available logarithmic interval.
To make this explicit, introduce
\begin{align}
& y=\cosh s,\quad \phi=\frac{u}{\sqrt{\sinh s}} \quad(q<q_e), \\
& y=\sinh t,\quad \phi=\frac{u}{\sqrt{\cosh t}} \quad(q>q_e).
\end{align}
Equation~\eqref{eq:bfv} then reduces to Eqs.~\eqref{eq:schro}.
Both potentials vanish exponentially at large $|s|$ or $|t|$, leaving the same asymptotic wave number $\nu$. 
Defining $\hat L\equiv\ln(2M/\sigma)$, the nonextremal interval then has length $\hat L+O(1)$, whereas the RN-like overcharged interval spans $2\hat L+O(1)$. 
Finite $O(1)$ endpoint corrections shift only the phase offsets.

The corresponding phase conditions are
\begin{align}
& \frac{\nu\hat L_n}{\pi}=n+n_0^-,\qquad q<q_e, \label{eq:qsub}\\
& \frac{2\nu\hat L_n}{\pi}=n+n_0^+,\qquad q>q_e.\label{eq:qover}
\end{align}
The offsets $n_0^\pm$ encode the finite matching phases, whereas the coefficient of $n$ determines the asymptotic spacing. 
Inverting these relations gives
\begin{align}
\sigma_n&\simeq2M\exp\!\left[-\frac{(n+n_0^-)\pi}{\nu}\right],\qquad q<q_e,
\label{eq:sigmasub}\\[2pt]
\sigma_n&\simeq2M\exp\!\left[-\frac{(n+n_0^+)\pi}{2\nu}\right],\qquad q>q_e,
\label{eq:sigmaover}
\end{align}
and hence
\begin{equation}
\frac{\sigma_{n+1}}{\sigma_n}\simeq e^{-\pi/\nu}
\quad (q<q_e),
\qquad
\frac{\sigma_{n+1}}{\sigma_n}\simeq e^{-\pi/(2\nu)}
\quad (q>q_e).
\label{eq:ratios}
\end{equation}
The doubled logarithmic interval on the overcharged side therefore accounts for the factor of two in the exponent.
Since $\nu\propto\sqrt{\alpha-\alpha_c}$, the generated scales are exponentially small in $1/\sqrt{\alpha-\alpha_c}$, giving the Miransky/BKT-type behavior relevant here~\cite{Miransky:1984ef}.

\paragraph{Near-extremal RN zero modes.}
For the existence curves, $\alpha$ plays the role of the eigenvalue.
Equation~\eqref{eq:qsub} gives the leading accumulation law
\begin{equation}
\alpha_n(q)-\alpha_c\simeq
\alpha_c\frac{4\pi^2(n+n_0^-)^2}{\hat L^2},
\end{equation}
which accounts for both the common accumulation at $\alpha_c$ and the slow $1/\ln^2(1/\sigma)$ convergence shown in Fig.~1.

For RN, the full linear equation~\eqref{eq:linear} admits the exact $\ell=0$ solution~\cite{Herdeiro:2018wub}
\begin{equation}
\phi(r)=P_\lambda\!\left(1+\frac{2Q^2(r-r_+)}{r(r_+^2-Q^2)}\right),
\qquad
\lambda=-\frac12+i\nu,
\end{equation}
where $\nu=\sqrt{\alpha-1/4}$ for RN. 
The solution is regular at $r=r_+$. 
At infinity, the argument reduces to $M/\sigma$, and the decay condition becomes
\begin{equation}
P_{-1/2+i\nu}\left(\frac{M}{\sigma}\right)=0.
\label{eq:RNzero}
\end{equation}
For $M/\sigma\gg1$, the Legendre function has the asymptotic form
\begin{equation}
P_\lambda(z)\simeq2|C|(2z)^{-1/2}\cos\!\left[\nu\ln(2z)+\arg C\right],
\qquad
C=\frac{\Gamma(i\nu)}{\sqrt\pi\,\Gamma(\tfrac12+i\nu)}.
\end{equation}
Using $\Gamma(i\nu)=\Gamma(1+i\nu)/(i\nu)$, $\arg\Gamma(1+i\nu)=-\gamma\nu+O(\nu^3)$, and $\arg\Gamma(\tfrac12+i\nu)=-(\gamma+2\ln2)\nu+O(\nu^3)$ gives $\arg C=-\pi/2+2\ln2\,\nu+O(\nu^3)$. 
The $n$th zero therefore satisfies $\nu\ln\frac{2M}{\sigma_n}+\arg C=\left(n+\frac12\right)\pi$,
which gives the phase offset
\begin{equation}
n_0^-=1-\frac{2\ln2}{\pi}\nu+O(\nu^2),
\label{eq:n0sub}
\end{equation}
and the onset scale
\begin{equation}
\sigma_n\simeq8M\exp\!\left[-\frac{(n+1)\pi}{\nu}\right].
\label{eq:refined}
\end{equation}
Thus both the constant phase shift and the prefactor $8M$ follow from the exact RN boundary-value problem. 
In particular, the factor $8=2e^{2\ln2}$ originates from the finite phase correction contained in the Legendre connection coefficient. 

Equation~\eqref{eq:n0sub} fixes the constant and linear terms of the nonextremal phase near the BF threshold, but the range displayed in Fig.~1 extends to $q_e-q=10^{-3}$, where finite-$\nu$ corrections are no longer negligible.
We therefore keep the analytically fixed terms in Eq.~\eqref{eq:n0sub} and parametrize the remaining correction as
\begin{equation}
n_0^-(\nu)
=
1-\frac{2\ln2}{\pi}\nu
+c_2\nu^2+c_3\nu^3+c_4\nu^4 .
\label{eq:n0fit}
\end{equation}
A single fit to all numerical eigenvalues gives $c_2=0.303762$, $c_3=-0.086217$, and $c_4=0.008358$.

\paragraph{Extremal scalarized sequence.}
On the overcharged extremal branch, Eq.~\eqref{eq:extrel} gives $\sigma=|Q_s|$. 
Equation~\eqref{eq:qover} then yields
\begin{equation}
\frac{|Q_{s,n}|}{M}\simeq 2e^{-(n+n_0^+)\pi/(2\nu)},\qquad
\frac{|Q_{s,n+1}|}{|Q_{s,n}|}\simeq e^{-\pi/(2\nu)}.
\label{eq:extscale}
\end{equation}
Unlike the RN zero-mode problem, the inner endpoint now lies in a nonlinear region where the scalar leaves the nonzero horizon stationary point. 
The throat equation therefore fixes the slope in $n$, while the finite phase $n_0^+$ is determined by matching to the full nonlinear solution. 

Taking the logarithm of the first relation in Eq.~\eqref{eq:extscale} gives $\ln(M/|Q_{s,n}|)=(n+n_0^+)\pi/(2\nu)-\ln2$. 
Since $1/(2\nu)=\sqrt{\alpha_c/(\alpha-\alpha_c)}$, this becomes
\begin{equation}
\ln\frac{M}{|Q_{s,n}|}=\pi(n+n_0^+)\sqrt{\frac{\alpha_c}{\alpha-\alpha_c}}+O(1).
\label{eq:bkt}
\end{equation}
The $O(1)$ term contains the explicit $-\ln2$ contribution together
with subleading corrections from the nonlinear matching.
Equation~\eqref{eq:bkt} therefore describes the Miransky scaling
inherited from the RN-like throat.

\end{document}